\documentclass[reprint,
superscriptaddress,
showpacs,preprintnumbers,
 amsmath,amssymb,
 aps,prl,
]{revtex4-1}

\usepackage{graphicx}
\usepackage{dcolumn}
\usepackage{bm}

\begin{document}

\title{Quantum Quench of an Atomic Mott Insulator}

\author{David Chen}
\author{Matthew White}
    \altaffiliation[Now at: ]{Stanford Research Systems Inc., 1290-D Reamwood Ave, Sunnyvale, CA 94089}
\author{Cecilia Borries}
    \altaffiliation[Now at: ]{Department of Atmospheric Sciences, University of Alaska, 930 Koyukuk Dr, Fairbanks, AK 99775}
\author{Brian DeMarco}
\affiliation{Department of Physics, University of Illinois, 1110 W Green St, Urbana, IL 61801}

\date{\today}

\begin{abstract}

We study quenches across the Bose-Hubbard Mott-insulator-to-superfluid quantum phase transition using an ultra-cold atomic gas trapped in an optical lattice.  Quenching from the Mott insulator to superfluid phase is accomplished by continuously tuning the ratio of Hubbard tunneling to interaction energy.  Excitations of the condensate formed after the quench are measured using time-of-flight imaging.  We observe that the degree of excitation is proportional to the fraction of atoms that cross the phase boundary, and that the quantity of excitations and energy produced during the quench have a power-law dependence on the quench rate.  These phenomena suggest an excitation process analogous to the Kibble-Zurek (KZ) mechanism for defect generation in non-equilibrium classical phase transitions.

\end{abstract}


\maketitle
The study of non-equilibrium phase transitions is wide ranging, touching on topics as diverse as the formation of structures in the early universe \cite{Kibble1976} and the practicality of adiabatic quantum computing \cite{PhysRevA.74.060304}.  The so-called ``Kibble-Zurek" (KZ) mechanism has been used to understand some universal features---principally the rate of topological defect formation---of quenches across classical phase transitions  \cite{Kibble1976,Zurek1985}.  ``Quench" in this context refers to rapidly varying a thermodynamic parameter in order to drive the system out of equilibrium.  The KZ theory has recently been extended to quantum phase transitions \cite{Cucchietti2007,Dziarmaga2010,Dziarmaga2008,DeGrandi2010,Grandi2010,Mathey2010,Mathey2009,Schir2011,Schutzhold2006,Snoek,Trefzger2010,PhysRevA.81.053604}. In contrast to the classical case, quantum phase transitions involve completely closed quantum mechanical evolution at zero temperature, for which quenches are accomplished by varying a parameter in the Hamiltonian in order to tune between different quantum phases. While the KZ mechanism has successfully been tested for classical transitions (e.g., on liquid crystals \cite{CHUANG15031991}), and spontaneous vortex formation has been observed during cooling an atomic gas through the Bose-Einstein condensation transition \cite{ISI:000260038300045}, experimental examination of quantum quenches have been scant.  Notably, there is evidence that the formation of ferromagnetic domains in a spin-1 Bose-Einstein condensate can be attributed to a sudden quantum quench \cite{ISI:000240622000040}.  In this work, we probe quantum quenches for a paradigm of quantum phase transitions---the Bose-Hubbard (BH) model---using atoms confined in an optical lattice.  In contrast to previous experiments \cite{ISI:000177788600033,Bakr30072010}, we quench from the Mott-insulator (MI) to the superfluid (SF) state, and we systematically investigate the formation of excitations as the quench amplitude and rate are varied.

In our experiment, a cubic optical lattice formed from three intersecting pairs of 812~nm laser beams is superimposed on a parabolically confined $^{87}$Rb Bose-Einstein condensate; details of our apparatus can be found in Ref. \cite{McKayPRA2009} and references therein.  The atoms in the lattice are described by the inhomogeneous BH model with tunneling energy $t$ and interaction energy $U$, the ratio of which is controlled by tuning the lattice laser intensity to adjust the lattice potential depth $s$.  By changing $s$,  MI and SF phases can be sampled inhomogeneously in the gas \cite{PhysRevA.71.063601}: for $s\gtrsim13 E_R$, nested Mott-insulator and superfluid layers exist in the lattice, and for $s\lesssim13 E_R$ the gas is purely superfluid, as shown in Fig. 1 ($E_R=h/8md^2$, where $m$ is the atomic mass, $d=406$~nm is the lattice spacing, and $h$ is Planck's constant).

Quenching across the SF--MI phase transition is accomplished by adjusting $s$ dynamically in such a way to transform the gas between equilibrium configurations with and without atoms in the MI phase present (Fig. 1).  While quenches are possible on all relevant timescales, in this paper we explore quenches that occur at rates $1/\tau_Q$ that are too slow to excite atoms into higher vibrational states in the lattice potential.  How $1/\tau_Q$ compares with the Hubbard energies is complicated because the phase boundary is crossed at a range of densities and $t/U$ in the trap, and therefore the Hubbard energies $U_c$ and $t_c$ at the phase transition change during the quench.  Despite this, the quenching rate is always slow compared with $U_c$: $1/\tau_Q$ varies from $1\times10^{-3}$--0.2~$U_c/h$. The quench rate is not consistently fast or slow compared with $t_c$ \footnote{For our fastest quenches, $1/\tau_Q$ is always fast compared with $t_c/h$ ($1/\tau_Q$:\;10--100\;$t_c/h$); for the slowest quenches, $1/\tau_Q$ is comparable to or slower than $t_c/h$ ($1/\tau_Q$:\;0.01--1\;$t_c/h$); and for intermediate speeds, $1/\tau_Q$ may be fast compared with $t_c/h$ at the beginning of the quench (e.g., at $s=25 E_R$) and slow at the end.} or the confining trap frequencies, the geometric mean of which varies from $43\pm2$~Hz at $s=0 E_R$ to $82\pm6$~Hz at $s=25 E_R$.

\begin{figure}[h]
\includegraphics[width=\linewidth]{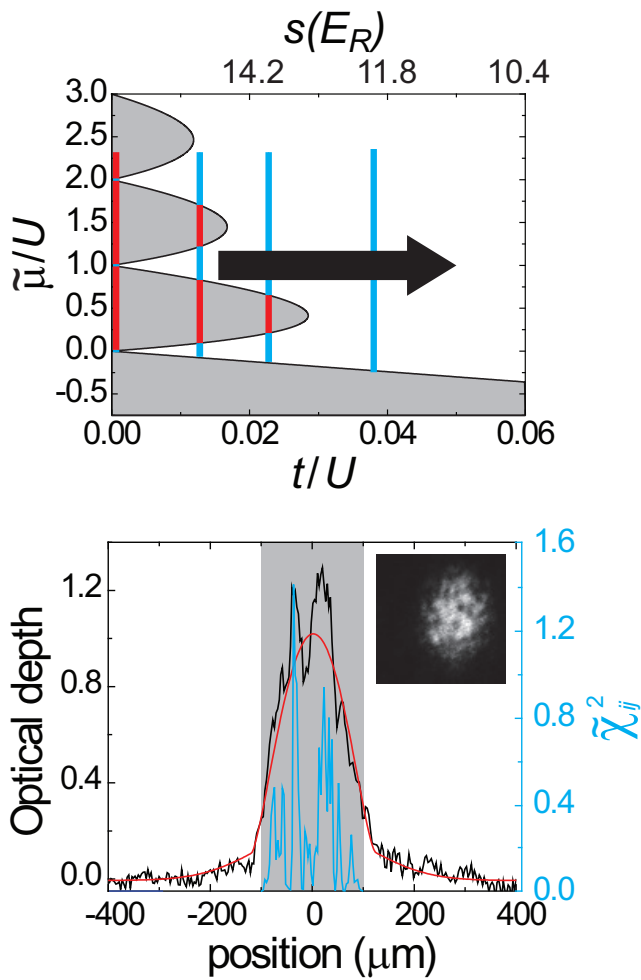}
\caption{\label{fig1} Quench across MI--SF boundary (top) and excitation measure (bottom).  As shown by the vertical lines, the trapped gas samples a range of densities and effective chemical potentials $\tilde{\mu}$ (in the LDA) \cite{PhysRevA.71.063601};  MI regions in the gas are colored red and SF blue. Given the overall confining potential in our experiment and atom number ($(161\pm13) \times 10^3$, averaged across all measurements), the maximum $\tilde{\mu}$ is roughly fixed at $2U$ for the measurements described here, which corresponds to a central filling of 4 atoms per site at low $s$ and a central MI with 3 atoms per site at high $s$.  A quench is accomplished by rapidly reducing  $s$ and thereby increasing $t/U$, as shown by the arrow.  A slice (black line) through a typical image (inset) taken after a quench for $s_0=25 E_R$ is displayed.  The image is fit to a smooth profile (red line), which is used to determine the deviation $\tilde{\chi}_{ij}^2$ (blue line) at each pixel in a masked region (gray).}
\end{figure}

The effect of variations in the fraction of atoms crossing the SF--MI phase boundary is investigated by quenching $s$ linearly in 5~ms to $s=4 E_R$ (corresponding to $t/U\approx1$, i.e., the purely SF regime), as shown in the inset to Fig 2 \footnote{A 5~ms total quench time was chosen for all data in this work so that the finite excitation (exponential) lifetime---measured to be $\approx30$~ms after a quench from $s_0=20E_R$---had little effect.}.  The ratio $t/U$ changes non-linearly during this quench; in the large $s$ limit, $t/U\propto e^{-2\sqrt{s}}$.  The fraction of atoms crossing the phase boundary is varied by adjusting the initial lattice potential depth $s_0$.  The data shown in Fig. 2 sample a range such that at high $s_0$ nearly all of the atoms start in the MI phase (with fillings ranging from 1--3 particles per site) and therefore cross the phase boundary; and at low enough $s_0$ so that all of the atoms are in the SF phase, and consequently no atoms cross the phase boundary.  The fraction of atoms in the MI phase before the quench, which is identical to the overall fraction of atoms traversing the phase boundary, is shown as a red line in Fig. 2, and is determined according to the measured atom number and a zero-temperature mean-field calculation in the LDA \cite{PhysRevA.71.063601}.  After the quench, the lattice is turned off in 200~$\mu$s.  This ``bandmapping" step \cite{McKayPRA2009}---which maps quasimomentum in the lattice to free momentum and suppresses atom diffraction---is necessary to improve the imaging signal-to-noise ratio given the long expansion times employed for these measurements.

The amount of excitation produced during the quench is determined by measuring the deviation from a smooth profile of time-of-flight images taken after release of the trapped gas immediately following bandmapping.  The absorption image is taken after a relatively long 50~ms of free expansion, so that vortices, if present, are visible \cite{PhysRevLett.84.806}, and phase gradients related to other topological or wave-like excitations are converted into large density fluctuations \cite{Dettmer2001,Petrov2001,Helwig2001}.  We fit the image to a smooth function $f$ that is a combination of a Thomas-Fermi profile and a Gaussian, and measure the amount of excitation $\tilde{\chi}^2$ as the deviation from the smooth profile: $\tilde{\chi}^2=\sum_{ij}\tilde{\chi}_{ij}^2=\alpha\sum_{ij}\frac{\left( OD_{ij}-f_{ij}\right)^2}{f_{ij}}/\sum_{ij}OD_{ij}$, where $i,j$ index the pixels in the image within a mask defined by an imaging signal-to-noise ratio greater than 5, $OD$ is the measured optical depth, and $\alpha$ is a proportionality constant that is determined using a numerical simulation. We find that all of the images used in this work are well described by this fit---a condensate appears present after the quench under all circumstances, and the condensate fraction varies from 0.35--0.6 across all of the data.  While it was suggested in Ref. \cite{PhysRevLett.89.250404} that the condensate fraction may oscillate after the quench, we find no evidence for such behavior.

The measure $\tilde{\chi}^2$ is chosen such that it is related to the fraction of atoms in excited states for the trapped, weakly interacting gas present before bandmapping.  The physical meaning of $\tilde{\chi}^2$ can be understood most straightforwardly for a one-dimensional non-interacting gas.  In this case, the density profile after sufficiently long TOF is the momentum distribution $n(q) = \left|\psi(q)\right|^2 = \left|\psi_0(q) + \delta \psi(q)\right|^2$, where $\delta\psi$ are plane-wave excitations, $\psi_0(q) = \sqrt{n_0(q)}$ is the  ground-state condensate wavefunction, and we work in the momentum representation. After averaging over random excitation phases, the number of atoms in excited states is $\int dq \: \left|\delta \psi(q)\right|^2 = \int dq \: \left[( n(q)-n_0(q) \right]^2/2 n_0(q)$. Given that $\int dq \:\: n_0(q)$ is the total number of atoms, $\tilde{\chi}^2$ is naturally interpreted as proportional to the number of excited atoms in the non-interacting limit.

Using a numerical simulation of the three-dimensional time-dependent Gross-Pitaevskii equation, we determined both that $\tilde{\chi}^2$ accurately reproduces the fraction of Bogoliubov excitations for a trapped condensate and the constant $\alpha$.  We start the simulation with a condensate at equilibrium in a parabolic trap (using the experimental atom number and trap parameters), and imprint Bogoliubov excitations under the LDA for a range of wavevectors corresponding to $0.8-3 \;\mu m^{-1}$; the Thomas-Fermi radius of the gas is approximately 10 $\mu$m before release. Images are generated by time evolving the condensate wavefunction for a free expansion and then integrating through the imaging line-of-sight.  The measure $\tilde{\chi}^2$ is determined for a range of excitation fractions averaged over 10 relative phases.  We determine that $\tilde{\chi}^2$ is equal to the fraction of excited atoms for $\alpha=10$ under simulated conditions.  This method is an approximation, and does not properly account for long-wavelength (i.e., trap-length-scale) excitations or topological excitations such as vortices, which are evident in the insets to Figs. 1 and 2.

As shown in Fig. 2, we find that the amount of excitation is proportional to the fraction of atoms crossing the phase boundary.  Below the emergence of the unit filling MI phase at $s_0\approx13 E_R$, $\tilde{\chi}^2$ is constant at $\tilde{\chi}^2_0\approx0.06$ (determined by averaging over all images with $s_0<13E_R$, and indicated by the dashed line in Fig. 2), a value that is consistent with the combination of photodetection shot noise and technical noise present in our imaging system.  Above $s_0\approx13 E_R$, the degree of excitation grows, until $\tilde{\chi}^2$ saturates to approximately 0.17 at high enough lattice depth, for which more than 90\% of the atoms are in the MI phase.

The behavior evident in Fig. 2 suggests that a Kibble-Zurek-like mechanism is responsible for generating excitations during the quench.  In the KZ picture, the diverging relaxation time near the phase boundary ``freezes in" fluctuations in the relative phase between atomic wavefunctions at different lattice sites present in the MI \cite{Dziarmaga2008}.  Some time after crossing the phase boundary, dynamics effectively restart, and the fluctuations develop into superfluid excitations, potentially including sound waves and topological excitations such as vortices. Given that only the regions of the lattice that cross the SF--MI phase boundary will give rise to excitations, the direct relation between the fraction initially in the MI phase and the degree of excitation is strong evidence for KZ physics.

\begin{figure}[h]
\includegraphics[width=\linewidth]{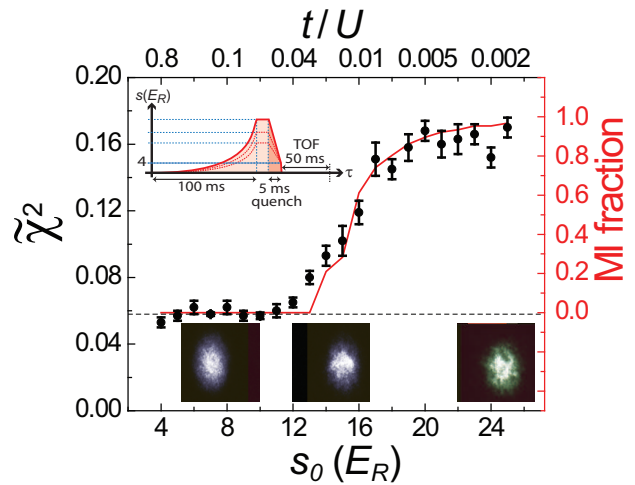}
\caption{\label{fig2} Amount of excitation produced by quenching as the fraction crossing the phase SF--MI boundary is varied.  The top left inset shows the experimental timeline; a magnetic field gradient is applied to support the atoms against gravity during TOF.  Characteristic TOF images are shown as insets for $s_0=$9, 15, and 25~$E_R$, and the error bars in this and the next figure are the standard deviation for the average taken over 5 images.  There is a 7\% systematic uncertainty to $s_0$, which is calibrated using Kapitza-Dirac diffraction.  The overall uncertainty in the MI fraction (red line) ranges from 30\% at $s_0=16 E_R$ to 10\% at $s_0=20 E_R$; below $s_0=12 E_R$ and above $s_0=22 E_R$ the uncertainty in MI fraction is zero.}
\end{figure}

In the KZ scenario, the quench rate controls the number of excitations generated according to a power law that depends on the critical exponents for the phase transition.  We measured this power law, as shown in Fig. 3, across two orders of magnitude in quench rate.  For this measurement we quench the lattice potential depth starting from a gas composed nearly entirely of the MI phase at $s_0=20 E_R$ (i.e., $t/U=0.005$) according to $s(\tau)=0.25 \ln^2\left(\frac{\pi a_s}{\sqrt{2} d}\frac{\tau}{\tau_Q}+e^{-2\sqrt{s_0}}\right)$ ($a_s\approx5$~nm is the scattering length) so that the rate of change $d(t/U)/d\tau\approx1/\tau_Q$ is approximately constant \cite{Zwerger}.  In addition to measuring how $\tilde{\chi}^2$ depends on $1/\tau_Q$, we also measure the kinetic energy generated by the quench.  The kinetic energy per particle $KE$ is measured from TOF images according to $KE= m \left\langle r^2 \right\rangle/2 \tau_{tof}^2$, with the second moment of the density distribution $\left\langle r^2 \right\rangle = 3/2 \cdot \sum_{ij} OD_{ij} r_{ij}^2/\sum_{ij}OD_{ij}$ after the TOF $\tau_{TOF}$.  The factor of 3/2 arises from assuming the energy is distributed equally among three directions.

\begin{figure}[h!]
\includegraphics[width=\linewidth]{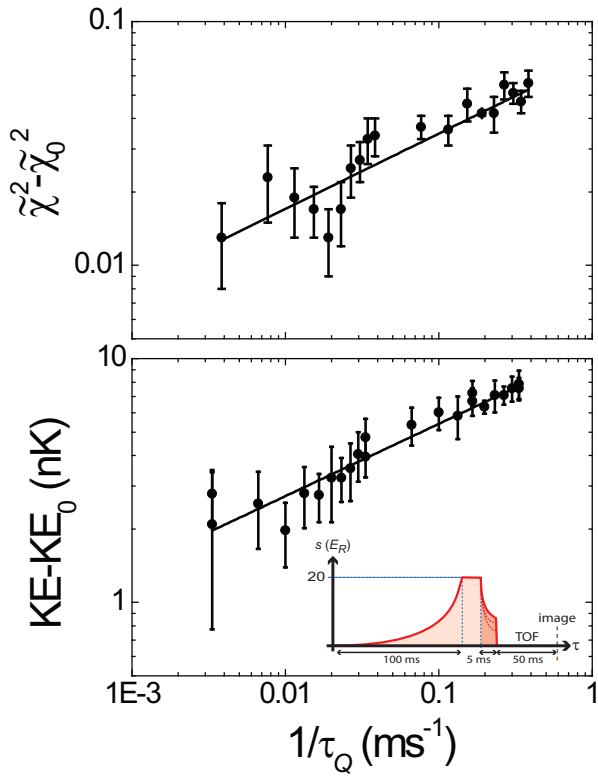}
\caption{\label{fig3} Excitation dependence on quench rate. The amount of excitation and kinetic energy determined from TOF images is shown as the quench rate $1/\tau_Q$ is varied. The measured offset $\tilde{\chi}^2_0$ is subtracted from $\tilde{\chi}^2$. Analogously, the measured expansion energy without the quench $KE_0$ (determined by averaging across images with $s_0<13E_R$) is subtracted from $KE$.  The inset shows the experimental timeline, and the solid lines are power-law fits to the data.}
\end{figure}

A fit to the data in Fig. 3 reveals power laws $1/\tau_Q^r$ for $\tilde{\chi}^2$ and $KE$ consistent within the fit uncertainty: $r=0.31\pm0.03$ and $0.32\pm0.02$, respectively.  While there are numerous detailed theoretical predictions for the number of excitations produced during a quench across the MI--SF phase transition \cite{Schutzhold2006,Snoek,Trefzger2010,Dziarmaga2008,Cucchietti2007}, none that we know of are directly applicable to our experiment. Generically, the size of domains associated with an excitation formed during the quench should scale as $\tau_Q^{\nu/(\nu z+1)}$ \cite{PhysRevB.72.161201}, where $z$ is the dynamical critical exponent and the correlation length diverges as $\xi\sim\left(t/U-t_c/U_c\right)^{-\nu}$ near the phase boundary.  In three dimensions, the density of excitations is therefore proportional to $1/\tau_Q^{3\nu/(\nu z+1)}$.  For our experiment, nearly all of the atoms cross the ``generic" phase transition and not the multi-critical point at the the ``tip" of the MI ``lobes."  In this case $\nu=1/2$ and $z=2,$ and therefore the number of excitations should scale as $1/\tau_Q^{3/4}$, which is inconsistent with our data.  This disagreement may be explained by numerous issues that deserve more theoretical attention.  For example, the spatially inhomogeneous nature of the phase transition that gives rise to a phase transition ``front" that moves through the gas has been examined in the context of the classical SF phase transition \cite{Retzker}, but not for the quantum case.  Also, the finite size of the gas will affect quench dynamics, as discussed in Ref. \cite{PhysRevA.81.053604} for the BH model in 1D.  Finally, since the data here were taken at low but finite temperature (the initial condensate fraction was more than 90\% before turning on the lattice), thermal effects may play an important role in the quench dynamics; see Ref. \cite{DeGrandi2010} for a analysis in the context of the Sine-Gordon model.

In conclusion, the method we have demonstrated provides a window into excited states and dynamics, which are beyond our current theoretical understanding in a wide variety of strongly interacting many-body quantum systems.  Quench dynamics may also have significant consequences for thermometry in optical lattice experiments \cite{McKay2010}.  One commonly employed technique to estimate temperature in a lattice is to slowly turn off the lattice potential, measure temperature, and then infer entropy in the lattice assuming that the turn off was adiabatic.  We find across a wide range of linear lattice quench rates that adiabaticity is violated; for example, for a quench from $s_0=20 E_R$, $\tilde{\chi}^2$ decreases from 0.17 to only 0.12 for turn off times varying from 5 to 25~ms.  Finally, in the future, optical lattice quenches may be investigated in other contexts, such as reduced dimensions and fermionic gases.

\begin{acknowledgments}
The authors acknowledge financial support from the National Science Foundation (award 0448354) and DARPA OLE program. Any opinions, findings, and conclusions or recommendations expressed in this material
are those of the authors and do not necessarily reflect the views of the National Science Foundation.
\end{acknowledgments}

\bibliography{quenching}

\end{document}